\documentclass[10pt,letterpaper]{article}
\usepackage[top=0.85in,left=2.75in,footskip=0.75in,marginparwidth=2in]{geometry}

\usepackage[utf8]{inputenc}
\usepackage{subcaption}
\usepackage{amsmath}
\usepackage{amssymb}
\usepackage{cite}
\usepackage[T1]{fontenc}

\usepackage{nameref,hyperref}

\usepackage[right]{lineno}

\usepackage{microtype}
\DisableLigatures[f]{encoding = *, family = * }

\raggedright
\usepackage{changepage}

\usepackage[aboveskip=1pt,labelfont=bf,labelsep=period,singlelinecheck=off]{caption}

\makeatletter
\renewcommand{\@biblabel}[1]{\quad#1.}
\makeatother

\usepackage{lastpage,fancyhdr,graphicx}
\usepackage{epstopdf}
\fancyheadoffset[L]{2.25in}
\fancyfootoffset[L]{2.25in}

\usepackage{color}

\definecolor{Gray}{gray}{.25}

\usepackage{graphicx}

\usepackage{sidecap}

\usepackage{wrapfig}
\usepackage[pscoord]{eso-pic}
\usepackage[fulladjust]{marginnote}
\reversemarginpar

\begin{document}
\vspace*{0.35in}

\begin{flushleft}
{\Large
\textbf\newline{More from Less? Environmental Rebound Effects of City Size}
}
\newline
\\
Joao Meirelles\textsuperscript{1*},
Fabiano Ribeiro\textsuperscript{2},
Gabriel Cury\textsuperscript,
Claudia Binder\textsuperscript{1}
\\
\bigskip
\bf{1} HERUS / EPFL
\\
\bf{2} DFI / UFLA
\\
\bigskip
* joao.meirelles@epfl.ch

\end{flushleft}

\section*{Abstract}
Global sustainability relies on our capacity of understanding and guiding urban systems, and their metabolism, in an adequate way. It has been proposed that bigger and denser cities are more resource-efficient than smaller ones because they tend to demand less infrastructure, consume less fuel for transportation and less energy for cooling / heating in per capita terms. This hypothesis is also called Brand's Law. But as cities get bigger, denser and more resource-efficient they also get richer. And richer inhabitants buy more, increasing its resource demand and associated environmental impacts. To fully understand the nexus between population size or density and the environmental impacts generated by a city, one needs to take into account both direct and indirect impacts. Facing the lack of empirical evidence on consumption-based emissions for cities, in this paper we propose a mean-field model to derive emissions estimations out of well-established urban metrics (city size, density, infrastructure, wealth). We aim at understanding if Brand's law holds true after adopting a consumption-based approach to urban emissions. The proposed model shows that when considering consumption-based emissions, in most cases Brand’s law falls apart - bigger cities have greater purchase power, resulting in greater consumption of goods and higher associated GHG emissions. The model also shows that decoupling between population and emissions is possible and dependent on the decoupling level between income and impacts. In order to achieve it, a shift in consumption patterns of most cities is of utmost importance, so that each new monetary unit added to the GDP, or any other income variable for that effect does not result in a proportional increase in GHG emissions.


\section{Introduction}

We live in a predominantly urban world, and rural to urban migration is not showing signs of slowing down. With population growth, the urban metabolism — all the exchanges between the city and the environment — is modified in diverse ways and we can expect challenges emerging in the management of, among others, environmental impacts, resource consumption, and waste disposal. Cities devour some 70\% of resources, and some 80\% of the energy consumed worldwide\cite{seto2015routledge}. It has become clear that global sustainability relies on our capacity of understanding and guiding urban systems, and their metabolism, in an adequate way.

\textbf{Strong is the faith in the dense city}. Urban density is well-established as a solution to sustainability in cities\cite{glaeser2010greenness, rybski2017cities} to the point of being identified by the IPCC as a crucial climate mitigation measure\cite{pachauri2014climate}. It is frequently argued that dense cities are more sustainable than sparse ones. As there is conclusive empirical evidence that density increases with population size \cite{angel2012planet,west2017scale,barthelemy2016structure}, it's been proposed that bigger cities are also more resource-efficient than smaller ones \cite{batty2013new}. This hypothesis - namely "As cities get bigger, they also get greener" - is called \textit{Brand's law} and stands for the fact that bigger and denser human settlements tend to demand less infrastructure per capita\cite{norman2006comparing}, less fuel consumption for transportation \cite{newman1989gasoline} and less energy consumption for cooling/heating\cite{behsh2002building}. Apparently, \textit{bigger cities do more with less} \cite{bettencourt2011bigger}. 

\textbf{"But a city is more than a place in space, it is a drama in time"}\cite{geddes1979civics}. Cities are not only the space they fill or the infrastructure they use for it. Cities are their social interactions as well. In fact, some will argue that cities are mainly defined by their social synergy\cite{bettencourt2013origins, jacobs1951life, soja2003writing}. These \textit{social reactors} accelerate economic activities and wealth creation\cite{bettencourt2007growth, jacobs2016economy}. As cities get bigger, they also get richer. And richer inhabitants buy more, increasing its resource demand and associated environmental impacts.

\begin{figure}[ht]
\centering
\includegraphics[width=0.958\linewidth]{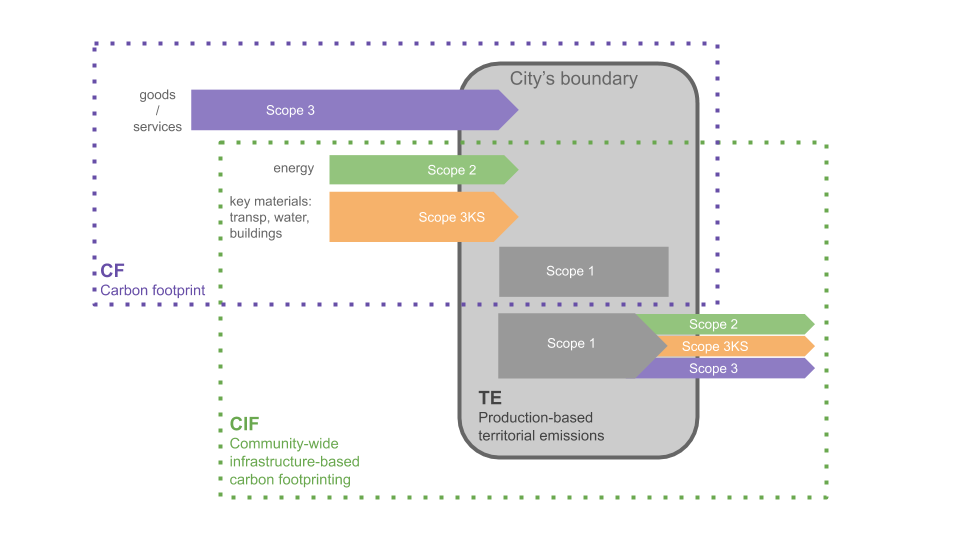}
\caption{Different emissions accounting methods and its relation to “Scope 1-3” concept. Based on \cite{chen2019review}}
\label{fig:ghg_methods}
\end{figure}

To fully understand the nexus between city size or density and the environmental impacts generated by a city, one needs to take into account both direct and indirect impacts. As urban agglomerations heavily rely on the supply of goods from outside their physical barriers, a consumption approach built on a life cycle perspective has been recommended by many specialists \cite{seto2014human, c40consumption, steininger2014justice}. For the case of urban greenhouse gas (GHG) emissions, there has been a gradual change from the first accounting protocols focusing on territorial production-based emissions to protocols accounting for upstream impacts of key materials and energy to very recently proposed protocols accounting for emissions embedded in upstream flows and goods consumed by the city\cite{chen2019review}. Figure \ref{fig:ghg_methods}, based on \cite{chen2019review}, visually describes the differences between GHG accounting methods commonly found in literature and their relation to the IPCC's "scope 1-3" concept\cite{ghg2001protocol, fong2015global}. While Territorial Emissions (TE) only accounts for emissions from sources within the city’s boundary (scope 1), Community-wide infrastructure-based carbon footprint (CIF) also includes emissions embodied in key materials like energy, water, and building materials (scope 2). Consumption-based carbon footprint (CF) adds to the previous emissions embodied in imports but deducts emissions embodied in exports. In this paper, we will adopt carbon footprint as the accounting method to understand the environmental impacts of cities.

Because of the novelty of the carbon footprint (CF) approach, the lack of data and complexity of the analysis, studies using a consumption-based approach generally i) focus on individual cases \cite{zhifu2014cons, dhakal2004urban}; and ii) rely on sample survey data and/or scaled-down data from national accounts \cite{minx2013carbon, gill2018ghg, heinonen2011dense, jones2014spatial, sudmant2018producer}. Up to now, findings indicate that consumption-based emission is substantially greater than production-based ones \cite{c40consumption}, and that denser/bigger cities have less direct emissions (energy / fuels locally consumed) but greater indirect emissions (embedded in goods and services) than smaller ones\cite{gill2018ghg, sudmant2018producer,heinonen2011dense,minx2013carbon, jones2014spatial}. However, due to the limited number of examples, the relationship between population size and total emissions (CF) remains unclear.

In the absence of assumption-free empirical evidence for multiple cities, our knowledge on environmental impacts of urban areas has not yet reached a systematic and mechanistic level. In this paper, we propose a theoretical approach based on the newly founded science of cities\cite{batty2013new}. During the last years, the science of cities has been established as a fertile quantitative approach to systematically understand the urban phenomena and the relationship between population, wealth, infrastructure and the size of human settlements based on urban data and models\cite{west2017scale, bettencourt2010unified}. One of its main pillars is the proposition that urban systems display universal scaling behavior regarding socioeconomic, infrastructural, and individual services\cite{bettencourt2007growth}. During the last decade, several studies reported urban variables, say $Y$, systematically scaling on a non-linear way with population $N$ for different urban systems \cite{bettencourt2007growth, bettencourt2016urban,ortman2014pre,gomez2012statistics,louf2014scaling,adhikari2017growth,meirelles2018evolution}. The relation between $Y$ and $N$ assumes the form $Y = Y_0 N^{\beta}$, where $Y_0$ is a constant and $\beta$ is the scaling exponent. The empirical findings suggest three different scaling regimes: superlinear $\beta>1$ for socioeconomic variables (e.g.: GDP, patents, AIDS cases), linear $\beta=1$ for individual basic services (e.g.: number of households, household water consumption), and sublinear $\beta<1$ for infrastructural variables (e.g.: total street length, number of schools). The observed pattern indicates that as cities get bigger, they foster human interactions, producing more social outputs with less infrastructural demand per capita. Some studies report deviations from the scaling hypothesis, indicating a high sensitivity of the exponent to the adopted definition of city \cite{arcaute2015constructing,louf2014smog}, the adopted method to estimate the exponent\cite{leitao2016scaling}, or sensibility to external factors such as macroeconomic structures \cite{strano2016rich,meirelles2018evolution} or federal policies\cite{muller2017does}. In this paper, we assumed the scaling hypothesis to be universal, given the amount and diversity of evidence supporting it. 

If we observe a lack of data on the explicit relation between city size and its environmental impacts, scientific evidence on the relation between city size, infrastructure demand, and wealth creation seems to be more robust. In this paper, we propose a mean-field model to derive emissions estimations out of well-established urban metrics (city size, density, infrastructure, wealth). We expect the stylized model to enable us to better understand the relationship between population size and environmental impacts in cities. To do so, we articulate both aspects of the problem at a broad theoretical level, a toy model. We want to understand if Brand's law holds true after adopting a consumption-based approach to urban emissions.

\section{On the Relation Between City Size and Environmental Impacts}
In this section, we will present empirical findings on the relations between city size, population density, income, infrastructure, and impacts. We will mostly use evidence coming from urban scaling literature. The stylized facts found in such relations will be further used in the construction or validation of our model. 

\subsection{City size and density}
Some studies used urban scaling theory to relate population size and urban density through the scaling exponent of area, normally found to be smaller than 1. Actually, urban scaling theory \cite{bettencourt2013origins,batty2011defining} predicts a scaling exponent of $\frac{2}{3}$ between population and land area. Works presenting empirical evidences found sublinear exponents between different area definitions and population size: for the global north Bettencourt \cite{bettencourt2016urban} found values between 0.85 and 0.95 for four European urban systems (UK, Italy, Germany, and France) as well as for Europe as a single Urban System; for the global south, Adhikari and Beurs\cite{adhikari2017growth} found scaling exponents between 0.5 and 0.8 in different years for African cities, while Meirelles et al. \cite{meirelles2018evolution} found an exponent of 0.84 for the Brazilian urban system; regarding ancient urban systems, Cesaretti \cite{cesaretti2016population} analyzed 173 medieval European settlements and found exponents close to $\frac{5}{6}$, while Ortman \cite{ortman2014pre} found values close to $\frac{2}{3}$ and $\frac{5}{6}$ for pre-Hispanic settlements in the Americas. There are still other works reporting sublinear scaling exponents between urban area and population \cite{stewart1958physics, nordbeck1971urban, paulsen2012yet}. Others studies, however, found counterexamples: Bettencourt \cite{bettencourt2016urban} found that the urbanized area scaled superlinearly with population in Spain; Cottineau \cite{cottineau2017diverse} found the exponent to be sensitive depending on the definition of city adopted in the scaling, with most values ranging in the sublinear spectra, but some revealed to be linear or superlinear. This sensitivity has also been analyzed and evidenced by Arcaute \cite{arcaute2015constructing}.

The diversity of papers, spamming through developed and developing countries and analyzing ancient as well as modern urban systems indicate that the population-density relation is likely to be universal to cities and is not related to contemporary dynamics or local specificities. Considering both pieces of evidence supporting and defying this claim, the great majority of findings seems to support the proposition that bigger cities are denser. This proposition is also supported by other methods \cite{angel2012planet}.

\subsection{City size and income}
Economic growth happens almost naturally with urbanization through agglomeration effects. Clustered people and businesses experience i) a higher probability of interactions and ii) the recombination of diverse capabilities since more people in a relative area are more likely to present different necessities, abilities and ideas \cite{fragkias2015urbanization}. These alone are already a source of economic growth \cite{jacobs2016economy, quigley1998urban}, and by reducing the distance between people, agglomeration also allows energy and time savings \cite{soja2003writing, fujita1989urban} which foment those interactions. This implies that as cities get bigger, they get richer.

Various empirical evidences support the claim that cities produce economic development and wealth. Among the urban scaling publications, many authors have shown superlinear relations between population and GDP, with an expected exponent of 1.15 or $\frac{7}{6}$ \cite{bettencourt2013origins, ribeiro2017model}. Empirical evidence has been found for China \cite{bettencourt2007growth, bettencourt2013origins}, Europe and European countries \cite{bettencourt2016urban, arcaute2015constructing, van2016urban}, Brazil \cite{meirelles2018evolution, alves2013distance} and the United States \cite{bettencourt2010unified}, among others, corroborating the superlinear relationship. Other economics-related variables also scale superlinearly with population, such as different aspects of income \cite{bettencourt2013origins, alves2013distance} wages \cite{bettencourt2007growth, ignazzi2015coevolution}, and expenditure \cite{sobolevsky2016cities}. On the other hand, Strano \cite{strano2016rich} found that although for European low-income cities the gross metropolitan product (GMP) scales superlinearly with population, for high-income cities it scales linearly, suggesting that from a certain stage of development, the GDP starts growing proportionally to the population. Arcaute \cite{arcaute2015constructing} found similar results for 535 cities in the UK and  Bettencourt \cite{bettencourt2010unified} for personal income in the USA. Once again, the majority of empirical evidence supports the proposition that on average, bigger cities are richer.

\subsection{City size and infrastructure}
Urban infrastructure tends to get more efficient as the city grows, and therefore, its variables in general scale sublinearly with population size. They are expected to scale with an exponent 0.85 or $\frac{5}{6}$ \cite{bettencourt2013origins},  originally proposed to be so by empirical evidence found for the length of electrical cables and road surface in Germany \cite{bettencourt2007growth}. Other studies found similar results for other urban systems: Kuhnert \cite{kuhnert2006scaling} for the length of low-voltage cables in Germany and the number of petrol stations in France, Germany, the Netherlands and Spain; Meirelles \cite{meirelles2018evolution} for the length of street and water supply networks and the number of primary and secondary schools in Brazil; Bettencourt \cite{bettencourt2013origins} for impervious surfaces in the world and in the EU, built area in China, area of roads in the USA and Germany and length of pipes in Japan; among others. This data point towards universality of urban infrastructure variables scaling sublinearly with population. To what concerns this paper, this relation indicates that bigger cities should have smaller material demand per capita for infrastructure.

However, infrastructure variables vary a great deal. The heterogeneity of processes governing such variables makes it a bit harder to assess their impacts categorically since they function and grow in significantly different ways. Following the scaling literature, they may be categorized into urban infrastructure (or public, such as the number of gas stations, hospitals, total length of water and gas pipelines, roads, etc.) and household infrastructure (or individual basic services, such as number of houses, number of bathrooms or bedrooms in houses, etc.). Studies have shown that infrastructure related to individual needs tend to scale linearly with population \cite{bettencourt2007growth, arcaute2015constructing, alves2013distance, meirelles2018evolution}. In spite of that, Schalpfer \cite{schlapfer2015urban} found that average building heights increase with population size, leading to a smaller surface-to-volume ratio in bigger cities, which implies that even if the number of household scales linearly with population, the resulting material and energy demand and their related environmental impacts probably increase sublinearly.

\subsection{City size and emissions}

Up to now, there’s no scientific consensus on the scaling regime of CF with population. In fact, most urban scaling studies consider only production-based emissions due to data availability, hindering our capacity to analyze the total Carbon Footprint and its relationship with population size. Fragkias \cite{fragkias2013does} conducted several different analyses along a decade of data for nearly a thousand core-based statistical areas in the US (366 MSAs and 576 Micropolitan Areas) and found almost-linear relationships in every case, with coefficients ranging from 0.9 to 0.95. The authors conclude that such findings refute the hypothesis that urban systems function similarly to biological ones, where the efficiency-size relationship is remarkably sublinear. Oliveira \cite{oliveira2014large} used remote sensing data to estimate production-based emissions from gridded data using 2281 clusters developed by a City Clustering Algorithm over the entire US territory and found that CO2 emissions scaled superlinearly with population with an average coefficient of 1.46. Even though the study did not consider consumption-based emissions, these results indicate that as population increases, cities become less efficient. Bettencourt \cite{bettencourt2016urban} also found a superlinear relation between CO2 emissions and population for 102 European Metropolitan areas with an exponent of 1.12, close to the value predicted by scaling theory of $\frac{7}{6}$. Whatever the method of assessment and estimation of CO2 emissions is not clear, and it is not possible to assume whether the emissions are consumption or production-based. Gudipudi \cite{gudipudi2019efficient} used four different sources for CO2 emissions data, mostly estimations based on production. On top of that, a small number of urban agglomerations from different countries were mixed to produce their results, indicating a lack of geographic consistency (scale analysis are expected to be performed within one single urban system). They found a superlinear relationship between CO2 emissions and population size for cities in Non-Annex I countries ($\beta$ = 1.18) and a sublinear relationship for cities in Annex I countries ($\beta$ = 0.87). 

Regardless of the results, all scaling studies lack consistent and harmonized consumption-based CO2 emissions data to properly analyze the cities’ Carbon Footprint, being the emissions usually estimated from production-based data alone. Problems emerging from data are closely related to methodological issues, making it impossible to reach a solid conclusion on how the total CO2 emissions scale with population and therefore, whether bigger cities are greener or not. A production-based figure is unable to differentiate between reduction and outsourcing of emissions in cities. Furthermore, the definition of city to be considered in each scaling greatly influences the resulting exponent. Louf \cite{louf2014smog} demonstrated that adopting two different definitions of cities by the Census Bureau in the US resulted in completely different scaling results (from $\beta$ = 0.95 to $\beta$ = 1.37, implying that larger cities can be greener or less green, respectively), an idea previously depicted for other urban variables \cite{arcaute2015constructing, cottineau2017diverse}. Gudipudi \cite{gudipudiurban} also points out two other significant problems in scaling data, the spatial resolution, and the regression method adopted, which can also change the results of a study. In short, the methodological inconsistencies among urban emission scaling studies, lack of data on consumption-based GHG emissions and the absence of a solid theoretical background prevent these conclusions from being extrapolated to other urban realities or creating general theoretical assumptions.

Studies analyzing the relation between urbanization and consumption-based emissions are performed with methodologies other than scaling. The majority of these studies - carried out in cities in developed countries - lead to the conclusion that direct emissions tend to decrease with greater populations, but indirect emissions tend to increase with greater populations. The relation between total emissions and population is defined by the intensity of direct emissions savings and indirect emissions increase and is country-dependent. Pang (forthcoming) analyzed the difference in emission patterns between rural and urban areas in Switzerland and found that direct emissions from urban areas were around 20\% smaller than from rural ones, while indirect emissions were from 5-10\% higher making the total emission in urban areas smaller than in rural ones. Gill \cite{gill2018ghg} analyzed agglomerations in Germany classified by population ranges and found that the bigger the population, the lower the direct emissions per capita and the greater the indirect emissions, leading to a slight decrease of total emissions in bigger cities. Minx \cite{minx2013carbon} found that scope 1 and 2 emissions in England decreased from rural to urban areas, while scope 3 increased, leading to constant total emissions per capita across the rural-to-urban range. Wiedenhofer \cite{wiedenhofer2013energy} found that urban households require less direct energy, but their total consumption is higher in Australia due to a big increase in indirect energy consumption. Heinonen \cite{heinonen2013situated} found that emissions from housing and transportation decreased for bigger and denser settlements in Finland, but indirect emissions more than overcome that trend leading to greater overall emissions. 

The latest report by the IPCC \cite{revi2014urban} states that, although limited, evidence suggests a higher consumption-based emission pattern in bigger cities than in rural areas for non-Annex I countries, with inconclusive results for Annex I countries - urban areas can have greater or lower consumption-based emissions when compared to rural settlements. Furthermore, considering energy-related emissions (Scope 2), the results are different for cities in Annex I and non-Annex I countries. Whilst two-thirds of the first show a lower per capita final energy use in comparison to their national averages, over two-thirds of the latter show the opposite result, indicating that the stage of development of the urban system greatly influences its emission patterns. Properly exploring this data and well-defining production and consumption-based emissions are of fundamental importance in order to reach a solid conclusion to adequately analyze the total CF of urban cities in comparison to rural areas and their national averages.

\begin{figure}[ht]
\captionsetup[subfigure]{justification=centering}
\begin{subfigure}{0.49\textwidth}
\includegraphics[width=0.95\linewidth]{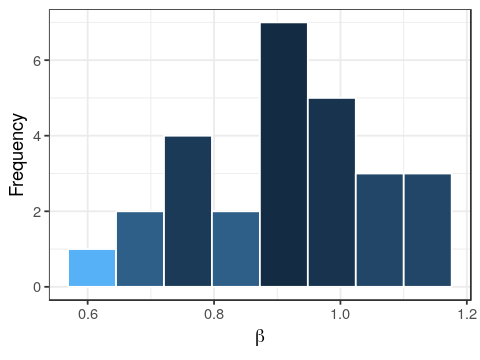} 
\label{fig:empirical_scaling}
\end{subfigure}
\begin{subfigure}{0.49\textwidth}
\includegraphics[width=0.95\linewidth]{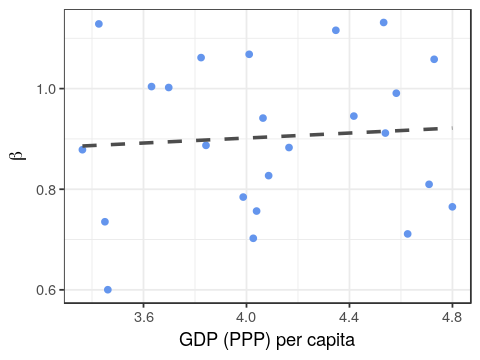}
\label{fig:beta_gdp}
\end{subfigure}
 
\caption{Scaling exponents ($\beta$) between population and carbon footprint for 27 countries from \cite{moran2018carbon}. Histogram (left) and relationship with GDP (PPP) per capita.}
\label{fig:moran}
\end{figure}

A recent study estimated consumption-based emissions (carbon footprints) for 13'000 cities across the globe (Gridded Global Model of City Footprints - GGMCF) \cite{moran2018carbon}.  To reach such estimations, authors had down-scaled national CFs and added existing sub-national CF studies. The process of down-scaling assumes that household expenditures profiles are constant for different city-size. This is unlikely to be the case and invalidates any study trying to understand the effect of city size on CF using the database. Figure \ref{fig:moran} (left) present a histogram of scaling exponent ($\beta$) between population and carbon footprint for 27 countries using GGMCF data \cite{moran2018carbon} and the relationship between ($\beta$) and the GDP of such countries. We estimated $\beta$ for all the countries with more than three cities in the GGMCF database. Differently from other variables, $\beta$ don't seem to follow a specific scaling behavior. It assumes sublinear, linear, and superlinear exponents for different countries, and the value of $\beta$ is not related to the wealth of the country (Figure \ref{fig:moran}-right). It is hard to assure the validity of such findings, given the down-scaling approach adopted in the model.

\subsection{Causal Diagram}

\begin{figure}[ht]
\centering
\includegraphics[width=0.958\linewidth]{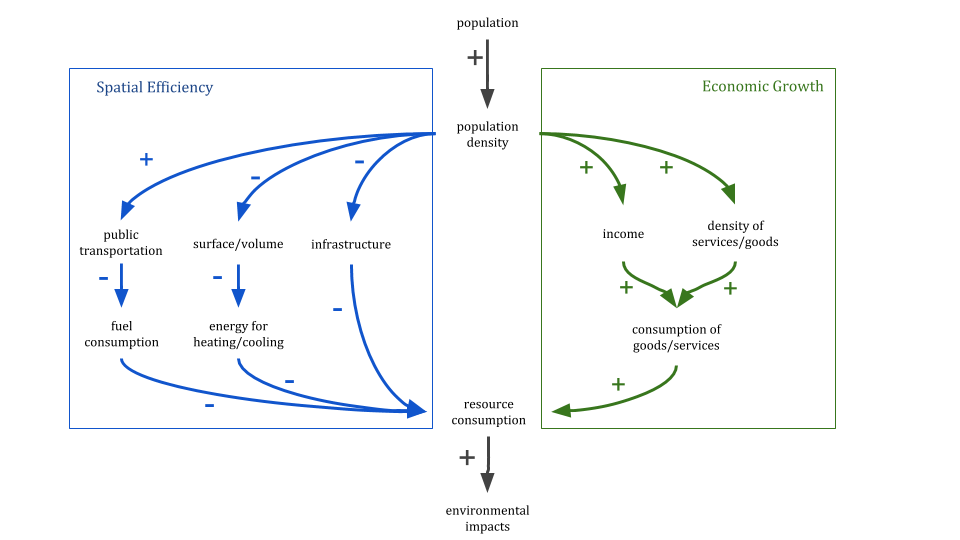}
\caption{Causal diagram between population size and environmental impact in cities}
\label{fig:cd}
\end{figure}

Figure \ref{fig:cd} summarizes the above described relations in a causal diagram. The diagram illustrates how population increase can influence on several variables which, in turn, result in higher or lower environmental impacts in the form of GHG emissions in a finite-size system. The methodologies commonly used focus on measuring variables from scope 1 and 2 emissions, related to direct emissions and, normally ruled by spatial efficiencies (blue box in the diagram). Scaling literature suggests that emissions related to fuel consumption, final energy use, and physical stock infrastructure should scale sublinearly with population in general, although some findings deviate from the expected regime. On the other hand, scope 3 emissions are expected to be driven by economic growth (green box in the diagram), where higher income and provision of goods and services lead to greater consumption and consequently, higher per capita GHG emissions. From a general point of view, we should expect a superlinear scaling regime with population for such emissions. The only way to adequately measure the Carbon Footprint of a city and understand its' relation to city size is by quantifying both types, including all three scopes. Our main goal is to derive a theory of scaling between consumption-based carbon footprint and city-size analyzing the balance between direct and indirect emissions in a theoretical standardized urban system.

This system structure, with opposing causalities (some reducing and others increasing the environmental impacts from population growth/density) resembles a well-known paradox in sustainability science: rebound effect. A rebound effect is the reduction of environmental gains from increasing efficiency of resource use due to systemic responses (behavioral, economic...)\cite{binswanger2001technological}. Imagine a household that implements house insulation saves energy and money and end up spending the saved money on a plane ticket to a far-away location at the end of the year. The final emissions balance of the insulation is likely to be positive if we consider the plane ticket. Situations like this, when the rebound exceeds the savings, are known as backfire rebound \cite{gillingham2016rebound}. Stimulating density as a way to achieve urban sustainability ignores rebound effects emerging from economic growth, another expected outcome of density. The type of sustainability discussion taking place in urban scaling literature tends to rest on resource-efficiency concepts ("Bigger cities do more with less" \cite{bettencourt2007growth, bettencourt2011bigger}), overestimating environmental savings by ignoring systemic responses stimulated by economic growth. In the following sections, we will implement a model, base on the scaling hypothesis, that account for such rebound effects.


\section{A toy model of urban carbon footprint based on scaling laws}
\subsection{Environmental rebound effect of city size}

\begin{figure}[ht]
\centering
\includegraphics[width=0.8\linewidth]{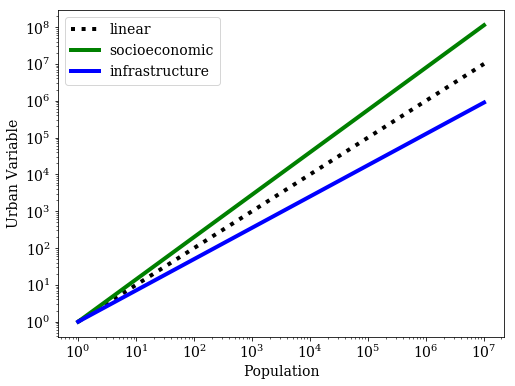}
\caption{Scaling of socioeconomic (green) and infrastructural (blue) variables in a theoretical mean field urban system, following expected scaling exponents. Black dotted line represents a linear scaling with the population.}
\label{fig:scaling_base}
\end{figure}

As a first approach, we propose a naive scaling model assuming scaling properties from literature. First of all, let us define the scaling properties we will use throughout the model. Over the last decade, urban scaling studies \cite{bettencourt2010unified} proposed that cities present universal and non-linear scaling between urban variables $Y$ and the population size $N$ of cities following the form $Y = Y_0 N^{\beta}$, where $Y_0$ is a constant and $\beta$ is the scaling exponent. Empirical evidence and theoretical models indicate that urban socioeconomic variables deriving from social interactions (e.g., GDP) scale in a superlinear ($\beta$ = 1.15) manner with population, while infrastructural variables scale sublinearly ($\beta$ = 0.85). Figure \ref{fig:scaling_base} represents the theoretical scaling relation of socioeconomic and infrastructural variables for cities with populations ranging from 1 to 10,000,000 inhabitants.

We'll assume that the \textit{total carbon footprint} of a city (in tons of CO$_2$) represented by  $C_{tot}$ can be divided into i) direct emissions, $C_{dir}$; and ii) indirect emissions, $C_{ind}$. That is, 

\begin{equation} \label{eq:1}
C_{tot} = C_{dir} + C_{ind}.
\end{equation}

We define $C_{dir}$ as the sum of territorial emissions (TE) and Community-wide infrastructure-based carbon footprint (CIF) that stay within the city boundaries. That is, it accounts for emissions occurring within the city boundary (e.g., gasoline consumption), emissions embedded in grid-supplied electricity, and emissions embedded in infrastructure (roads, water pipes, buildings), but not for emissions occurring in factories or other production processes. Differently, $C_{ind}$ is defined here as all other emissions taking place outside the city boundaries as a consequence of activities occurring within the city boundaries (e.g., embedded emissions in goods and services). 

Emissions from a given activity can be estimated by multiplying the activity data, which can be computed in monetary terms, by the activity emission factor $\overline{\epsilon}$ \cite{fong2015global}. In our toy model, we will assume that $\overline{\epsilon}$ can be described as the mean emission factor for the type of activity (direct or indirect). If every activity (e.g., buying a t-shirt, or 1 kg of cement, or 1 kWh of electricity) holds one associated emission factor $\overline{\epsilon}$, one can draw a probability distribution of emission factors by money expenditure for a city and estimate its average value.

In our model, $C_{dir}$ describes emissions related to spatially-embedded infrastructures, such as upstream emissions from resources for the building stocks, embedded emissions in electricity for housing and in gasoline for transportation. Let's consider that $C_{dir}$ of a city is directly dependent on its total expenditure in infrastructure. In this way, one can write that  $C_{dir} = \overline{\epsilon}_{dir} \cdot Y_{inf}$, where $Y_{inf}$ is the city’s total expenditure related to spatially-embedded infrastructure (in purchasing power parities (PPP) measured in National currency units/US dollar). And $\overline{\epsilon}_{dir}$ is the \textit{mean direct emission factor}.

On the other hand, $C_{ind}$ describes emissions related to the consumption of goods and services by the population and it's driven by the population's income. In our model, we will assume that $C_{ind}$ is directly dependent on the city’s total income (in PPP USD), which will be represented by $Y_{inc}$. Then, one can write that  $C_{ind} = \overline{\epsilon}_{ind}\cdot Y_{inc}$, where $\overline{\epsilon}_{ind}$ is the
\textit{mean indirect emission factor}. With a better definition of $C_{dir}$ and $C_{ind}$, Equation \ref{eq:1} can be re-written as

\begin{equation} \label{eq:2}
C_{tot} =  \overline{\epsilon}_{dir} \cdot Y_{inf} + \overline{\epsilon}_{ind} \cdot Y_{inc}.
\end{equation}

Here we bring in the stylized facts observed in urban scaling hypothesis. We will assume that the city’s total expenditure related to spatially-embedded infrastructure scales as any infrastructure-related variable, following the power-law 

\begin{equation} \label{eq:4}
Y_{inf} \approx Y_{inf}^0 N^{\beta_{inf}},
\end{equation}
where $\beta_{inf}<1$ (sublinear regime), according to the  empirical evidence.

Similarly, we will assume, based on the urban scaling hypothesis, that the city’s disposable income follow the expected scaling behavior of socioeconomic variables, known to scale in a super-linear manner with the city's population $N$ by the power-law 

\begin{equation} \label{eq:3}
Y_{inc} \approx Y_{inc}^0 N^{\beta_{inc}},
\end{equation}
where $Y_{inc}^0$ is the intercept, and  $\beta_{inc}$ is the scaling exponent. This is an approximation: disposable income is just a fraction of the city's GDP and although other variables related to expenditure, income, and wages had been observed to scale in a superlinear manner with population \cite{sobolevsky2016cities,bettencourt2013origins, alves2013distance,bettencourt2007growth, ignazzi2015coevolution}, to the best of our knowledge no study had explicitly estimated the scaling of disposable income.

Empirical evidence suggests $\beta_{inc}>1$ (superlinear regime). If both scaling exponents are correlated\cite{bettencourt2013origins, ribeiro2017model, west2017scale}, that is if  $\beta_{inc} = 1 + \delta$ and $\beta_{inf} = 1 - \delta$, then Equation \ref{eq:2} can be re-written as follows

\begin{equation} \label{eq:5}
C_{tot} = \overline{\epsilon}_{dir} \cdot Y_{inf}^0 N^{1 - \delta} + \overline{\epsilon}_{ind} \cdot Y_{inc}^0 N^{1 + \delta}.
\end{equation}

\begin{figure}[ht]
    \centering
    \includegraphics[width=0.8\linewidth]{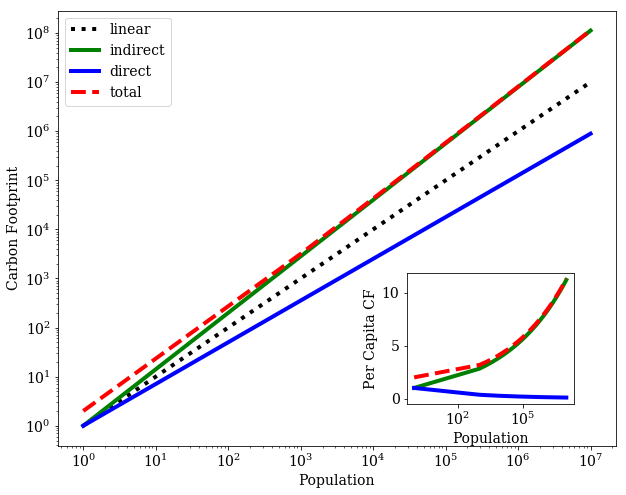}
    \caption{Scaling of Carbon Footprint in cities - Naive Scaling. Embedded in the bottom-right is the \textit{per capita} CF against the city population. $\overline{\epsilon}_{dir}= \overline{\epsilon}_{ind}= Y_{inf}^0= Y_{inc}^0 = 1$ and $\delta = 0.15$}
    \label{fig:naive}
\end{figure}

By solving equation~\ref{eq:5} computationally, we got the results presented in Figure \ref{fig:naive}. In order to understand the naive scaling of CF, we assumed 
$\overline{\epsilon}_{dir}= \overline{\epsilon}_{ind}= Y_{inf}^0= Y_{inc}^0 = 1$ and $\delta = 0.15$. Later on, we will further explore the effect of each variable.

It is possible to observe that, as the population $N$ increases, the indirect carbon footprint $C_{ind}$ (green line) becomes more and more important to the total carbon emission  $C_{tot}$ (red dashed line). That is, when $N$ is sufficiently large, $C_{tot} \sim  N^{1+\delta}$. This means a super-linear power-law relation between the total carbon emission and the population size, and it is independent of the values of the emission factors and the intercepts. The naive scaling model contradicts Brand's law or the belief that bigger cities are greener. The naive scaling model indicates a backfire rebound effect from city size/density: once emissions embedded in goods and services (scope 3) are included on top of the spatial-related emissions, bigger cities are expected to have higher emissions \textit{per capita} than smaller ones (Figure \ref{fig:naive} bottom-right). We will see below that it is possible to find certain configurations of parameters that maintain the scaling sublinear for a limited but still feasible population range (for example 1,000 $\leq$ N  $\leq$ 10,000,000). 

\begin{figure}[ht]
\captionsetup[subfigure]{justification=centering}
\begin{subfigure}{0.49\textwidth}
\includegraphics[width=0.99\linewidth]{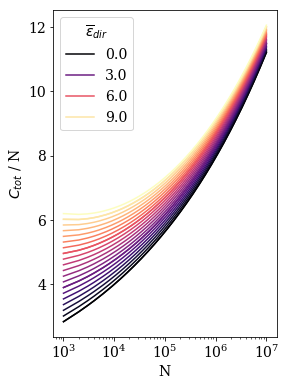} 
\caption{Direct Emissions Factor}
\label{fig:ef_dir}
\end{subfigure}
\begin{subfigure}{0.49\textwidth}
\includegraphics[width=0.99\linewidth]{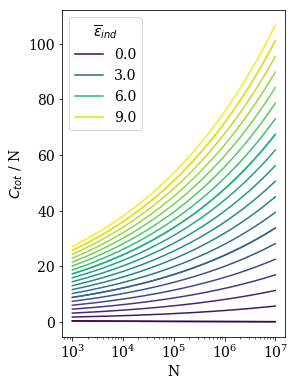}
\caption{Indirect Emissions Factor}
\label{fig:ef_ind}
\end{subfigure}
 
\caption{Per capita Carbon Footprint for varying emission factors}
\label{fig:image4}
\end{figure}

The Naive model assumes the same emission factor for both direct and indirect emissions ($\overline{\epsilon}_{dir}$=$\overline{\epsilon}_{ind})$) and the same intercept for both the income and infrastructure ($Y_{inf}^0 = Y_{inc}^0$). This is unlikely to be realistic, and in this subsection, we explore the effect of varying emission factors and intercepts on the scaling of emissions (\ref{fig:image4}). 

While the direct emissions factor ($\overline{\epsilon}_{dir}$) has only a small effect on the slope of the curve describing the relation between the population $N$ of a city and the \textit{per capita} carbon footprint $C_{tot}$, the indirect emission factor ($\overline{\epsilon}_{ind}$) can completely decouple emission from the city size. For very small indirect emission factors, bigger cities don't present greater emissions. Different values of $\overline{\epsilon}_{ind}$ diverge the trajectory of \textit{per capita} CF, while different values of $\overline{\epsilon}_{dir}$ produce very similar curves of \textit{per capita} CF. This happens because, for very small indirect emission factors, the superlinear scaling of indirect CF can be neglected. However, this only happens for very small values of $\overline{\epsilon}_{ind}$, which might be an unreal scenario only achievable by very efficient production processes or by an impact-less consumption pattern by most of the city's population. The same analysis holds true for the intercepts, with $Y_{inc}^0$ producing a similar behavior as $\overline{\epsilon}_{ind}$.

As a matter of fact, equation \ref{eq:5} can be re-written as follows.

\begin{eqnarray*} \label{eq:6}
C_{tot} &=& \overline{\epsilon}_{dir} \cdot Y_{inf}^0 N^{1 - \delta} + \overline{\epsilon}_{ind} \cdot Y_{inc}^0 N^{1 + \delta}\\
&=& N [\overline{\epsilon}_{dir} \cdot Y_{inf}^0 N^{- \delta} + \overline{\epsilon}_{ind} \cdot Y_{inc}^0 N^{+ \delta}]\\
&=& N \overline{\epsilon}_{dir} \cdot Y_{inf}^0 \left[ N^{- \delta} + \frac{\overline{\epsilon}_{ind} \cdot Y_{inc}^0}{\overline{\epsilon}_{dir} \cdot Y_{inf}^0} N^{+ \delta}\right]\\
&\therefore&\\
C_{tot}&\propto& N  \left[ N^{- \delta} + \left(
\frac{\overline{\epsilon}_{ind}}{\overline{\epsilon}_{dir}} \right)
\left( \frac{Y_{inc}^0}{Y_{inf}^0}\right) N^{\delta}\right]\\
\end{eqnarray*}

Which can be re-written in \textit{per capita} terms

\begin{equation} \label{eq:7}
\frac{C_{tot}}{N} \propto   N^{-\delta} + \left(
\frac{\overline{\epsilon}_{ind}}{\overline{\epsilon}_{dir}} \right)
\left( \frac{Y_{inc}^0}{Y_{inf}^0}
\right) N^{\delta}
\end{equation}

\begin{figure}[ht]
\centering
\includegraphics[width=0.8\linewidth]{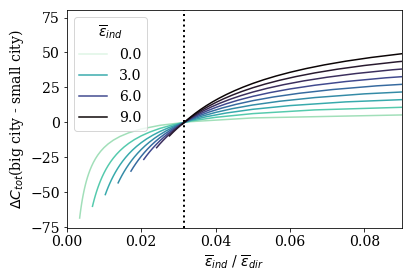}
\caption{Difference between \textit{per capita} $C_{tot}$ of the biggest city and the \textit{per capita} $C_{tot}$ of the smallest city in the model ($\Delta$ $C_{tot}$(${big city}$ -${small city}$)) as a function of $\overline{\epsilon}_{ind}/\overline{\epsilon}_{dir}$. The inflexion point is independent of the absolute value of $\overline{\epsilon}_{ind}$ (represented by different line colors in Fig. 6). The same behaviour is true for $Y_{inf}^0/Y_{inc}^0$. Values smaller than zero mean that smaller cities have greater emissions than bigger cities, while values greater than zero means that smaller cities have smaller emissions than bigger ones.}
\label{fig:parameter_matrix}
\end{figure}

This results show how the ratios between the emission factors ($\overline{\epsilon}_{ind}/\overline{\epsilon}_{dir}$) and the intercepts 
($Y_{inc}^0/Y_{inf}^0$) influence the dynamics of \textit{per capita} CF. When one of the factors assume very small values $N^{-\delta}$ govern the equation, producing a sublinear scaling. When the indirect emissions factor is much smaller than the direct emissions factor ($\overline{\epsilon}_{ind} \ll \overline{\epsilon}_{dir}$) or the income intercept is much smaller than the infrastructure spend intercept ($Y_{inc}^0 \ll Y_{inf}^0$), bigger cities become greener than smaller ones. Figure \ref{fig:parameter_matrix} present per capita emissions difference between the smallest (N = 1,000) and biggest (N = 10,000,000) cities in the simulation as a function of $\overline{\epsilon}_{ind}/\overline{\epsilon}_{dir}$. It is possible to observe that for $\overline{\epsilon}_{ind}/\overline{\epsilon}_{dir}$ < 0.03 bigger cities emit less carbon \textit{per capita} than smaller ones, while for  $\overline{\epsilon}_{ind}/\overline{\epsilon}_{dir}$ > 0.03 bigger cities emit more carbon \textit{per capita} than smaller ones, regardless of the absolute parameter values. The same behaviour is true for $Y_{inf}^0/Y_{inc}^0$.

\subsection{The role of Decoupling}
There is a hidden assumption in the previous model: 

\begin{equation} \label{eq:8}
\frac{\partial \overline{\epsilon}}{ \partial N} \sim 0 \therefore \frac{\partial \overline{\epsilon}}{ \partial Income} \sim 0
\end{equation}

This means that the impact factors are constant across the population range of simulated cities. In other words the naive version assumes that i)for direct emissions: the materials and production processes adopted in infrastructure construction and operation are the same across cities with different populations and wealth levels and ii)for indirect emissions: citizens with different income levels and from cities of all sizes consume the same type of goods and services, following the same consumption distribution. This is very unlikely to be true, given that i)bigger cities are able to implement different infrastructure technologies than smaller ones, with different environmental impacts \cite{schlapfer2015urban} and ii)within the same urban system, inhabitants of bigger cities tend to be wealthier than those from smaller cities, and wealthier people do have different consumption patterns and associated environmental impacts than poor ones \cite{hubacek2017global}. It is important to test under which conditions bigger cities are greener than smaller ones considering all these effects. 

What we will be testing is if decoupling can offset the rebound effect of city size. Decoupling is defined as the reduction of environmental impacts per unit of economic output, meaning that the economy can grow with a less than proportional increase in it's associated environmental impacts (relative decoupling) or even without increasing at all it's associated environmental impacts (absolute decoupling) \cite{international2011decoupling}. Coupling is also possible: when an economy experiences an increase in environmental impacts per unit of economic output. Consumption-based decoupling are relatively rare to observe, but it can happen in countries across all levels of economic development: a recent study \cite{akizu2018decoupling} found that between 2000-2014, out of 124 countries, 27 have experienced absolute decoupling, while 17 faced relative decoupling and 80 have endured a coupling between carbon footprint and economic growth. While absolute decoupling for the whole economy might be impossible to sustain indefinitely \cite{ward2016decoupling}, it can be the case that a relative decoupling together with the spatial efficiency of cities retrieves Brand's law for an urban system. 

To reach a more realistic model and test the role of decoupling in urban sustainability, we introduced a correction factor to equation \ref{eq:5}: the emission factor will be considered a function of wealth, that is

\begin{equation} \label{eq:9}
C_{tot} = \overline{\epsilon}_{dir}(Y_{inf}) \cdot Y_{inf} + \overline{\epsilon}_{ind}(Y_{inc}) \cdot Y_{inf}.
\end{equation}

We can infer the expected relation between income and emissions factor ($\overline{\epsilon}$) from the better-studied relation between income and carbon emissions ($C_{tot}$), known to follow a power law with an exponent ($\gamma$) \cite{chakravarty2009sharing, brown2011energetic, hubacek2017global}. That is

\begin{equation}\label{eq:10}
C = C^0 Y^{\gamma},    
\end{equation}
which allows us to write

\begin{eqnarray*} \label{eq:11}
\overline{\epsilon}(Y)\propto\frac{C}{Y}&=& C^0 Y^{\gamma-1} \\
\overline{\epsilon}(Y)&=& \rho Y^{\gamma-1}. \\
\end{eqnarray*}

Which, assuming that wealth (Y) also scales as a power law with the population (N), can be re-written as 

\begin{equation} \label{eq:12}
\overline{\epsilon}(Y)= \rho N^{(1+\delta).(\gamma-1)}
\end{equation}

Here we propose that direct and indirect emissions would assume different scaling exponents

\begin{equation} \label{eq:13}
\overline{\epsilon}_{dir}(Y_{inf}) = C_{dir}^0 N^{(1+\delta).(\gamma_{inf}-1)}
\end{equation}

\begin{equation} \label{eq:14}
\overline{\epsilon}_{ind}(Y_{inc}) = C_{ind}^0 N^{(1+\delta).(\gamma_{inc}-1)}
\end{equation}

Which relate to the fact that the carbon intensity of infrastructure and consumption follow different decoupling behaviors. While the first is mainly driven by technology and investment capacity, the second is also driven by social norms.

We can now re-rite equation \ref{eq:9} as follow

\begin{equation} \label{eq:15}
C_{tot} =  A \cdot N^{(1+\delta).(\gamma_{inf}-1)} \cdot N^{1 - \delta} + B \cdot N^{(1+\delta).(1-\gamma_{inc})} \cdot N^{1 + \delta}
\end{equation}

\begin{equation} \label{eq:16}
C_{tot} =  A \cdot N^{\gamma_{inf}(1-\delta)} + B \cdot N^{\gamma_{inc}(1+\delta)}
\end{equation}

\begin{figure}[ht]
\captionsetup[subfigure]{justification=centering}
\begin{subfigure}{0.49\textwidth}
\includegraphics[width=0.99\linewidth]{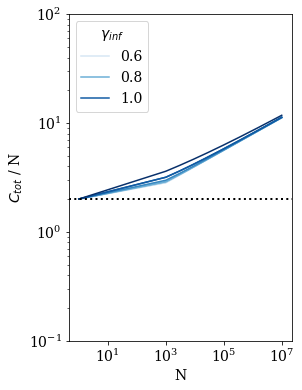} 
\caption{$\gamma_{inf}$}
\label{fig:gamma_dir}
\end{subfigure}
\begin{subfigure}{0.49\textwidth}
\includegraphics[width=0.99\linewidth]{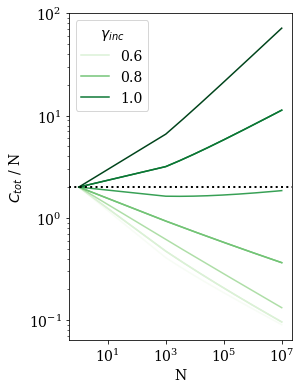}
\caption{$\gamma_{inc}$}
\label{fig:gamma_ind}
\end{subfigure}
 
\caption{Per capita Carbon Footprint for different values for of $\gamma$}
\label{fig:image7}
\end{figure}

Empirical values for $\gamma$ vary significantly from country to country. Authors tend to agree that most common values in country-wide regressions are smaller than 1 \cite{chakravarty2009sharing, brown2011energetic, hubacek2017global}, but a recent consumption-based study found values evenly distributed between 0.5 and 1.5 for regression within countries \cite{moran2018carbon}. It's worth noticing that, once again, empirical evidence here is mainly based on production accounting methods at the national level. Our strategy to deal with such empirical faults will be to test the "phase-space" of the scaling exponent, instead of proposing one value. Figure \ref{fig:image7} presents the scaling between \textit{per capita} $C_{tot}$ and city population in our model, considering different values for both $\gamma_{inf}$ and $\gamma_{inc}$. We assumed $A=C_{dir}^0 \cdot Y_{inf}^0$ and $B=C_{ind}^0 \cdot Y_{inc}^0$. 

It is possible to observe that $\gamma_{inc}$, which describes the elasticity between income an indirect emissions is the key variable in the model which define the scaling regime between population and carbon footprint. With a small value of $\gamma_{inc}$, very different results can be obtained: values above 0.8 produces superlinear scaling of \textit{per capita} $C_{tot}$ with population, while values below 0.8 produce sublinear scaling. Notice that $\gamma=1$ retrieve the naive model, with $\overline{\epsilon}$ assuming a constant relation with income and population. This relation could explain the diversity of relations found between city size and consumption-based emissions for different countries. On top of that, the fact that countries from the global south tend to present higher values for $\gamma$ \cite{akizu2018decoupling} can potentially explain the predominance of the conclusion "bigger is not greener" in those countries, while smaller values for $\gamma$ in the global south could lead to "bigger is greener", as empirically observed \cite{revi2014urban}.

In order to analytically find the critical value of $\gamma$ that shifts the scaling regime of $C_{tot}$ from superlinear to sublinear ($\gamma*$), we can rewrite equation \ref{eq:15} as follow:

\begin{eqnarray*} \label{eq:17}
C_{tot} &=&  A \cdot N^{\gamma} \cdot N^{-\gamma \cdot \delta} + B \cdot N^{\gamma} \cdot N^{\gamma \cdot \delta} \\
C_{tot} &=&  N^{\gamma} [A \cdot N^{-\gamma \cdot \delta} + B \cdot N^{\gamma \cdot \delta}]\\
\end{eqnarray*}

Letting $N \xrightarrow{}\infty$, $B \cdot N^{\gamma \cdot \delta} \xrightarrow{}0$ 
\begin{equation} \label{eq:18}
\therefore C_{tot} \approx  N^{\gamma(1+\delta)}
\end{equation}

Therefore, the limit between superlinear and sublinear scaling is given by
\begin{equation} \label{eq:19}
\gamma^* (1+\delta) = 1 \therefore \gamma^* = \frac{1}{1+\delta}
\end{equation}

\begin{figure}[ht]
\centering
\includegraphics[width=0.8\linewidth]{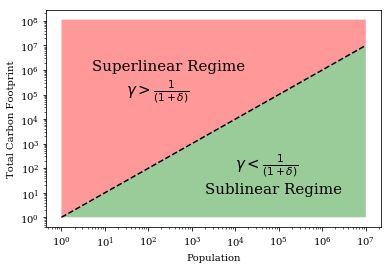}
\caption{Analytical phase-space diagram of expected scaling regime of total carbon footprint ($C_{tot}$) given different values of $\gamma$. Red regions indicates a superlinear scaling between population and $C_{tot}$, while green regions indicates a sublinear scaling regime. Dashed line represents $\gamma^*=\frac{1}{(1+\delta)}$}
\label{fig:gramma_critico}
\end{figure}

Figure \ref{fig:gramma_critico} illustrate the phase-space diagram of expected scaling regime of total carbon footprint ($C_{tot}$) given different values of $\gamma$. For $\gamma>\frac{1}{(1+\delta)}$, total emissions scale in a superlinear manner with population,  while for $\gamma<\frac{1}{(1+\delta)}$ a sublinear regime is observed. For $\gamma=\frac{1}{(1+\delta)}$ a linear regime is retrieved, with total emissions growing proportionally to the population ($\gamma^*$ represented by the dashed line in figure \ref{fig:gramma_critico})

\section{Discussion}
The proposed model suggests that Brand's Law's premise - bigger cities are greener - is not entirely false. Nevertheless, it carries along a gross methodological error: our model shows that if we consider only territorial-based emissions, then yes, bigger cities are almost always greener. However, our model also shows that when considering consumption-based emissions, in most cases Brand’s law falls apart - bigger cities have greater purchase power, resulting in greater consumption of goods and higher associated GHG emissions. An even weaker conclusion can be reached considering the final Carbon Footprint since it involves the already inaccurate measures of both direct and indirect emissions. 

It must also be highlighted that the significance of indirect consumption-based emissions is underestimated, with most studies on urban sustainability considering only the direct emissions when proposing conclusions and emissions patterns\cite{gudipudi2019efficient,rybski2017cities,ribeiro2019effects}. Its estimation is harder and less direct, and there is not a universal methodology of estimation. On top of everything, it introduces a policy problem: since the emissions due to consumption are generated elsewhere, in places out of the reach of the cities institutions, the implementation of mitigation policies is hindered. A global effort in transboundary mitigation policies is necessary in order to reduce indirect emissions.

All things considered, the model also shows that decoupling between population and emissions is possible and dependent on the decoupling level between income and impacts. In order to achieve it, a shift in consumption patterns of most cities is of utmost importance, so that each new monetary unit added to the GDP, or any other income variable for that effect does not result in a proportional increase in GHG emissions.

The model still carries along some gross assumptions due to the lack of empirical evidences: i) the scaling exponent of disposable income is proportional to the scaling exponent of total GDP in a system of cities; ii)Prices in cities across the population interval are the same, there is no price differences between a big city and a small city. The presented model could be fine-tuned by i)including scaling exponent explicitly related to disposable income of cities; ii)including a city-cite parity price model. Those changes, however, should not dismiss the general findings from the model. Although the exact value of $\gamma^*$ for which bigger cities start to be "greener" would probably change, the fact that there is a limit-value should not change.

From a broader perspective the model proposed in this work is a mean field toy model, not accounting for local specificities but general standards, limiting the practical application of its results. A statistical mechanics approach should be developed in order to account for local specificities. Due to the lack of empirical evidence, the model could not be validated. One way to overcome this is by using credit card data associated with multi-regional input-output emission models (MRIO), generating emissions estimations from consumption data. Using the same methodology for a range of cities could allow a more reliable comparison between their indirect emissions estimations.






\section*{Acknowledgments}
We thank just about everybody.
Rao, Camilo, Vinicius, Franziska, Emanuele, Vincent, Melissa, ALberto, Pekka, Romano

\nolinenumbers

\bibliography{library}

\begin{thebibliography}{10}

\bibitem{seto2015routledge}
Karen~C Seto, William~D Solecki, and Corrie~A Griffith.
\newblock {\em The Routledge handbook of urbanization and global environmental
  change}.
\newblock Routledge, 2015.

\bibitem{glaeser2010greenness}
Edward~L Glaeser and Matthew~E Kahn.
\newblock The greenness of cities: Carbon dioxide emissions and urban
  development.
\newblock {\em Journal of urban economics}, 67(3):404--418, 2010.

\bibitem{rybski2017cities}
Diego Rybski, Dominik~E Reusser, Anna-Lena Winz, Christina Fichtner, Till
  Sterzel, and J{\"u}rgen~P Kropp.
\newblock Cities as nuclei of sustainability?
\newblock {\em Environment and Planning B: Urban Analytics and City Science},
  44(3):425--440, 2017.

\bibitem{pachauri2014climate}
Rajendra~K Pachauri, Myles~R Allen, Vicente~R Barros, John Broome, Wolfgang
  Cramer, Renate Christ, John~A Church, Leon Clarke, Qin Dahe, Purnamita
  Dasgupta, et~al.
\newblock {\em Climate change 2014: synthesis report. Contribution of Working
  Groups I, II and III to the fifth assessment report of the Intergovernmental
  Panel on Climate Change}.
\newblock IPCC, 2014.

\bibitem{angel2012planet}
Shlomo Angel.
\newblock {\em Planet of cities}.
\newblock Lincoln Institute of Land Policy Cambridge, MA, 2012.

\bibitem{west2017scale}
Geoffrey~B West.
\newblock {\em Scale: the universal laws of growth, innovation, sustainability,
  and the pace of life in organisms, cities, economies, and companies}.
\newblock Penguin, 2017.

\bibitem{barthelemy2016structure}
Marc Barthelemy.
\newblock {\em The structure and dynamics of cities}.
\newblock Cambridge University Press, 2016.

\bibitem{batty2013new}
Michael Batty.
\newblock {\em The new science of cities}.
\newblock MIT press, 2013.

\bibitem{norman2006comparing}
Jonathan Norman, Heather~L MacLean, and Christopher~A Kennedy.
\newblock Comparing high and low residential density: life-cycle analysis of
  energy use and greenhouse gas emissions.
\newblock {\em Journal of urban planning and development}, 132(1):10--21, 2006.

\bibitem{newman1989gasoline}
Peter~WG Newman and Jeffrey~R Kenworthy.
\newblock Gasoline consumption and cities: a comparison of us cities with a
  global survey.
\newblock {\em Journal of the American planning association}, 55(1):24--37,
  1989.

\bibitem{behsh2002building}
Basam Behsh.
\newblock Building form as an option for enhancing the indoor thermal
  conditions.
\newblock In {\em Build. Phys.-6th Nord. Symp}, 2002.

\bibitem{bettencourt2011bigger}
Lu{\'\i}s~MA Bettencourt and Geoffrey~B West.
\newblock Bigger cities do more with less.
\newblock {\em Scientific American}, 305(3):52--53, 2011.

\bibitem{geddes1979civics}
Patrick Geddes.
\newblock {\em Civics: as applied sociology}.
\newblock Leicester University Press, 1979.

\bibitem{bettencourt2013origins}
Lu{\'\i}s~MA Bettencourt.
\newblock The origins of scaling in cities.
\newblock {\em science}, 340(6139):1438--1441, 2013.

\bibitem{jacobs1951life}
Jane Jacobs.
\newblock {\em The death and life of great American cities}.
\newblock ., 1951.

\bibitem{soja2003writing}
Edward Soja.
\newblock Writing the city spatially.
\newblock {\em City}, 7(3):269--280, 2003.

\bibitem{bettencourt2007growth}
Lu{\'\i}s~MA Bettencourt, Jos{\'e} Lobo, Dirk Helbing, Christian K{\"u}hnert,
  and Geoffrey~B West.
\newblock Growth, innovation, scaling, and the pace of life in cities.
\newblock {\em Proceedings of the national academy of sciences},
  104(17):7301--7306, 2007.

\bibitem{jacobs2016economy}
Jane Jacobs.
\newblock {\em The economy of cities}.
\newblock Vintage, 2016.

\bibitem{seto2014human}
Karen~C Seto, Shobhakar Dhakal, Anthony Bigio, Hilda Blanco, Gian~Carlo
  Delgado, David Dewar, Luxin Huang, Atsushi Inaba, Arun Kansal, Shuaib Lwasa,
  et~al.
\newblock Human settlements, infrastructure and spatial planning.
\newblock {\em .}, 2014.

\bibitem{c40consumption}
C40.
\newblock {\em Consumption-based GHG emissions of C40 cities}.
\newblock c40, 2018.

\bibitem{steininger2014justice}
Karl Steininger, Christian Lininger, Susanne Droege, Dominic Roser, Luke
  Tomlinson, and Lukas Meyer.
\newblock Justice and cost effectiveness of consumption-based versus
  production-based approaches in the case of unilateral climate policies.
\newblock {\em Global Environmental Change}, 24:75--87, 2014.

\bibitem{chen2019review}
Guangwu Chen, Yuli Shan, Yuanchao Hu, Kangkang Tong, Thomas~Oliver Wiedmann,
  Anu Ramaswami, Dabo Guan, Lei Shi, and Yafei Wang.
\newblock A review on city-level carbon accounting.
\newblock {\em Environmental science \& technology}, 2019.

\bibitem{ghg2001protocol}
WBCSD and WRI.
\newblock {\em The greenhouse gas protocol: a corporate accounting and
  reporting standard}.
\newblock WBCSD: World Business Council for Sustainable Development; WRI: World
  Resources Institute; World Resources Inst, 2001.

\bibitem{fong2015global}
Wee~Kean Fong, Mary Sotos, M~Michael~Doust, Seth Schultz, Ana Marques, and
  Chang Deng-Beck.
\newblock Global protocol for community-scale greenhouse gas emission
  inventories (gpc).
\newblock {\em World Resources Institute: New York, NY, USA}, 2015.

\bibitem{zhifu2014cons}
Zhifu Mi, Yunkun Zhang, Dabo Guan, Yuli Shan, Zhu Liu, Ronggang Cong, Xiao-Chen
  Yuan, and Yi-Ming Wei.
\newblock Consumption-based emission accounting for chinese cities.
\newblock {\em Ecological Indicators}, 47:26--31, 2014.

\bibitem{dhakal2004urban}
Shobhakar Dhakal and Imura Hidefumi.
\newblock {\em Urban energy use and greenhouse gas emissions in Asian
  mega-cities: Policies for a sustainable future}.
\newblock Urban Environmental Management Project, Institute for Global
  Environmental~…, 2004.

\bibitem{minx2013carbon}
Jan Minx, Giovanni Baiocchi, Thomas Wiedmann, John Barrett, Felix Creutzig,
  Kuishuang Feng, Michael F{\"o}rster, Peter-Paul Pichler, Helga Weisz, and
  Klaus Hubacek.
\newblock Carbon footprints of cities and other human settlements in the uk.
\newblock {\em Environmental Research Letters}, 8(3):035039, 2013.

\bibitem{gill2018ghg}
Bernhard Gill and Simon Moeller.
\newblock Ghg emissions and the rural-urban divide. a carbon footprint analysis
  based on the german official income and expenditure survey.
\newblock {\em Ecological Economics}, 145:160--169, 2018.

\bibitem{heinonen2011dense}
Jukka Heinonen, Riikka Kyr{\"o}, and Seppo Junnila.
\newblock Dense downtown living more carbon intense due to higher consumption:
  a case study of helsinki.
\newblock {\em Environmental Research Letters}, 6(3):034034, 2011.

\bibitem{jones2014spatial}
Christopher Jones and Daniel~M Kammen.
\newblock Spatial distribution of us household carbon footprints reveals
  suburbanization undermines greenhouse gas benefits of urban population
  density.
\newblock {\em Environmental science \& technology}, 48(2):895--902, 2014.

\bibitem{sudmant2018producer}
Andrew Sudmant, Andy Gouldson, Joel Millward-Hopkins, Kate Scott, and John
  Barrett.
\newblock Producer cities and consumer cities: Using production-and
  consumption-based carbon accounts to guide climate action in china, the uk,
  and the us.
\newblock {\em Journal of cleaner production}, 176:654--662, 2018.

\bibitem{bettencourt2010unified}
Luis Bettencourt and Geoffrey West.
\newblock A unified theory of urban living.
\newblock {\em Nature}, 467(7318):912, 2010.

\bibitem{bettencourt2016urban}
Lu{\'\i}s~MA Bettencourt and Jos{\'e} Lobo.
\newblock Urban scaling in europe.
\newblock {\em Journal of The Royal Society Interface}, 13(116):20160005, 2016.

\bibitem{ortman2014pre}
Scott~G Ortman, Andrew~HF Cabaniss, Jennie~O Sturm, and Luis~MA Bettencourt.
\newblock The pre-history of urban scaling.
\newblock {\em PloS one}, 9(2):e87902, 2014.

\bibitem{gomez2012statistics}
Andres Gomez-Lievano, HyeJin Youn, and Luis~MA Bettencourt.
\newblock The statistics of urban scaling and their connection to zipf’s law.
\newblock {\em PloS one}, 7(7):e40393, 2012.

\bibitem{louf2014scaling}
R{\'e}mi Louf, Camille Roth, and Marc Barthelemy.
\newblock Scaling in transportation networks.
\newblock {\em PLoS One}, 9(7):e102007, 2014.

\bibitem{adhikari2017growth}
Pradeep Adhikari and Kirsten~M de~Beurs.
\newblock Growth in urban extent and allometric analysis of west african
  cities.
\newblock {\em Journal of land use science}, 12(2-3):105--124, 2017.

\bibitem{meirelles2018evolution}
Joao Meirelles, Camilo~Rodrigues Neto, Fernando~Fagundes Ferreira,
  Fabiano~Lemes Ribeiro, and Claudia~Rebeca Binder.
\newblock Evolution of urban scaling: Evidence from brazil.
\newblock {\em PloS one}, 13(10):e0204574, 2018.

\bibitem{arcaute2015constructing}
Elsa Arcaute, Erez Hatna, Peter Ferguson, Hyejin Youn, Anders Johansson, and
  Michael Batty.
\newblock Constructing cities, deconstructing scaling laws.
\newblock {\em Journal of The Royal Society Interface}, 12(102):20140745, 2015.

\bibitem{louf2014smog}
R{\'e}mi Louf and Marc Barthelemy.
\newblock Scaling: lost in the smog.
\newblock {\em Environment and Planning B: Planning and Design},
  41(5):767--769, 2014.

\bibitem{leitao2016scaling}
Jorge~C Leit{\~a}o, Jos{\'e}~Mar{\'\i}a Miotto, Martin Gerlach, and Eduardo~G
  Altmann.
\newblock Is this scaling nonlinear?
\newblock {\em Royal Society open science}, 3(7):150649, 2016.

\bibitem{strano2016rich}
Emanuele Strano and Vishal Sood.
\newblock Rich and poor cities in europe. an urban scaling approach to mapping
  the european economic transition.
\newblock {\em PloS one}, 11(8):e0159465, 2016.

\bibitem{muller2017does}
Nicholas~Z Muller and Akshaya Jha.
\newblock Does environmental policy affect scaling laws between population and
  pollution? evidence from american metropolitan areas.
\newblock {\em PloS one}, 12(8):e0181407, 2017.

\bibitem{batty2011defining}
Michael Batty and Peter Ferguson.
\newblock Defining city size, 2011.

\bibitem{cesaretti2016population}
Rudolf Cesaretti, Jos{\'e} Lobo, Lu{\'\i}s~MA Bettencourt, Scott~G Ortman, and
  Michael~E Smith.
\newblock Population-area relationship for medieval european cities.
\newblock {\em PloS one}, 11(10):e0162678, 2016.

\bibitem{stewart1958physics}
John~Q Stewart and William Warntz.
\newblock Physics of population distribution.
\newblock {\em Journal of regional science}, 1(1):99--121, 1958.

\bibitem{nordbeck1971urban}
Stig Nordbeck.
\newblock Urban allometric growth.
\newblock {\em Geografiska Annaler: Series B, Human Geography}, 53(1):54--67,
  1971.

\bibitem{paulsen2012yet}
Kurt Paulsen.
\newblock Yet even more evidence on the spatial size of cities: Urban spatial
  expansion in the us, 1980--2000.
\newblock {\em Regional Science and Urban Economics}, 42(4):561--568, 2012.

\bibitem{cottineau2017diverse}
Cl{\'e}mentine Cottineau, Erez Hatna, Elsa Arcaute, and Michael Batty.
\newblock Diverse cities or the systematic paradox of urban scaling laws.
\newblock {\em Computers, environment and urban systems}, 63:80--94, 2017.

\bibitem{fragkias2015urbanization}
Michail Fragkias.
\newblock Urbanization, economic growth and sustainability.
\newblock In {\em The Routledge Handbook of Urbanization and Global
  Environmental Change}, pages 33--50. Routledge, 2015.

\bibitem{quigley1998urban}
John~M Quigley.
\newblock Urban diversity and economic growth.
\newblock {\em Journal of Economic perspectives}, 12(2):127--138, 1998.

\bibitem{fujita1989urban}
Masahisa Fujita.
\newblock {\em Urban economic theory: land use and city size}.
\newblock Cambridge university press, 1989.

\bibitem{ribeiro2017model}
Fabiano~L Ribeiro, Joao Meirelles, Fernando~F Ferreira, and Camilo~Rodrigues
  Neto.
\newblock A model of urban scaling laws based on distance dependent
  interactions.
\newblock {\em Royal Society open science}, 4(3):160926, 2017.

\bibitem{van2016urban}
Anthony~FJ Van~Raan, Gerwin Van Der~Meulen, and Willem Goedhart.
\newblock Urban scaling of cities in the netherlands.
\newblock {\em PLoS One}, 11(1):e0146775, 2016.

\bibitem{alves2013distance}
Luiz~GA Alves, Haroldo~V Ribeiro, Ervin~K Lenzi, and Renio~S Mendes.
\newblock Distance to the scaling law: a useful approach for unveiling
  relationships between crime and urban metrics.
\newblock {\em Plos one}, 8(8):e69580, 2013.

\bibitem{ignazzi2015coevolution}
CA~Ignazzi.
\newblock {\em Coevolution in the Brazilian system of cities}.
\newblock PhD thesis, th{\`e}se de Doctorat, Universit{\'e} Paris 1,
  Panth{\'e}on-Sorbonne, Paris, France, 2015.

\bibitem{sobolevsky2016cities}
Stanislav Sobolevsky, Izabela Sitko, Remi~Tachet Des~Combes, Bartosz Hawelka,
  Juan~Murillo Arias, and Carlo Ratti.
\newblock Cities through the prism of people’s spending behavior.
\newblock {\em PloS one}, 11(2):e0146291, 2016.

\bibitem{kuhnert2006scaling}
Christian K{\"u}hnert, Dirk Helbing, and Geoffrey~B West.
\newblock Scaling laws in urban supply networks.
\newblock {\em Physica A: Statistical Mechanics and its Applications},
  363(1):96--103, 2006.

\bibitem{schlapfer2015urban}
Markus Schl{\"a}pfer, Joey Lee, and Lu{\'\i}s Bettencourt.
\newblock Urban skylines: building heights and shapes as measures of city size.
\newblock {\em arXiv preprint arXiv:1512.00946}, 2015.

\bibitem{fragkias2013does}
Michail Fragkias, Jos{\'e} Lobo, Deborah Strumsky, and Karen~C Seto.
\newblock Does size matter? scaling of co2 emissions and us urban areas.
\newblock {\em PLoS One}, 8(6):e64727, 2013.

\bibitem{oliveira2014large}
Erneson~A Oliveira, Jos{\'e}~S Andrade~Jr, and Hern{\'a}n~A Makse.
\newblock Large cities are less green.
\newblock {\em Scientific reports}, 4:4235, 2014.

\bibitem{gudipudi2019efficient}
Ramana Gudipudi, Diego Rybski, Matthias~KB L{\"u}deke, Bin Zhou, Zhu Liu, and
  J{\"u}rgen~P Kropp.
\newblock The efficient, the intensive, and the productive: Insights from urban
  kaya scaling.
\newblock {\em Applied energy}, 236:155--162, 2019.

\bibitem{gudipudiurban}
Ramana Gudipudi, Diego Rybski, Matthias~KB Ludeke, and Jurgen~P Kropp.
\newblock Urban emission scaling—research insights and a way forward.
\newblock {\em Environment and Planning B: Urban Analytics and City Science},
  2019.

\bibitem{wiedenhofer2013energy}
Dominik Wiedenhofer, Manfred Lenzen, and Julia~K Steinberger.
\newblock Energy requirements of consumption: Urban form, climatic and
  socio-economic factors, rebounds and their policy implications.
\newblock {\em Energy policy}, 63:696--707, 2013.

\bibitem{heinonen2013situated}
Jukka Heinonen, Mikko Jalas, Jouni~K Juntunen, Sanna Ala-Mantila, and Seppo
  Junnila.
\newblock Situated lifestyles: I. how lifestyles change along with the level of
  urbanization and what the greenhouse gas implications are—a study of
  finland.
\newblock {\em Environmental Research Letters}, 8(2):025003, 2013.

\bibitem{revi2014urban}
A~Revi, DE~Satterthwaite, F~Arag{\'o}n-Durand, J~Corfee-Morlot, RBR Kiunsi,
  M~Pelling, DC~Roberts, and W~Solecki.
\newblock Urban areas climate change 2014: Impacts, adaptation, and
  vulnerability. part a: Global and sectoral aspects. contribution of working
  group ii to the fifth assessment report of the intergovernmental panel on
  climate change ed cb field et al.
\newblock {\em Field, CB, Barros, VR, Dokken, DJ et al}, pages 535--612, 2014.

\bibitem{moran2018carbon}
Daniel Moran, Keiichiro Kanemoto, Magnus Jiborn, Richard Wood, Johannes
  T{\"o}bben, and Karen~C Seto.
\newblock Carbon footprints of 13 000 cities.
\newblock {\em Environmental Research Letters}, 13(6):064041, 2018.

\bibitem{binswanger2001technological}
Mathias Binswanger.
\newblock Technological progress and sustainable development: what about the
  rebound effect?
\newblock {\em Ecological economics}, 36(1):119--132, 2001.

\bibitem{gillingham2016rebound}
Kenneth Gillingham, David Rapson, and Gernot Wagner.
\newblock The rebound effect and energy efficiency policy.
\newblock {\em Review of Environmental Economics and Policy}, 10(1):68--88,
  2016.

\bibitem{hubacek2017global}
Klaus Hubacek, Giovanni Baiocchi, Kuishuang Feng, Ra{\'u}l~Mu{\~n}oz Castillo,
  Laixiang Sun, and Jinjun Xue.
\newblock Global carbon inequality.
\newblock {\em Energy, Ecology and Environment}, 2(6):361--369, 2017.

\bibitem{international2011decoupling}
International~Resource Panel, United Nations Environment Programme.~Sustainable
  Consumption, and Production Branch.
\newblock {\em Decoupling natural resource use and environmental impacts from
  economic growth}.
\newblock UNEP/Earthprint, 2011.

\bibitem{akizu2018decoupling}
Ortzi Akizu-Gardoki, Gorka Bueno, Thomas Wiedmann, Jose~Manuel Lopez-Guede,
  I{\~n}aki Arto, Patxi Hernandez, and Daniel Moran.
\newblock Decoupling between human development and energy consumption within
  footprint accounts.
\newblock {\em Journal of cleaner production}, 2018.

\bibitem{ward2016decoupling}
James~D Ward, Paul~C Sutton, Adrian~D Werner, Robert Costanza, Steve~H Mohr,
  and Craig~T Simmons.
\newblock Is decoupling gdp growth from environmental impact possible?
\newblock {\em PloS one}, 11(10):e0164733, 2016.

\bibitem{chakravarty2009sharing}
Shoibal Chakravarty, Ananth Chikkatur, Heleen De~Coninck, Stephen Pacala,
  Robert Socolow, and Massimo Tavoni.
\newblock Sharing global co2 emission reductions among one billion high
  emitters.
\newblock {\em Proceedings of the National Academy of Sciences},
  106(29):11884--11888, 2009.

\bibitem{brown2011energetic}
James~H Brown, William~R Burnside, Ana~D Davidson, John~P DeLong, William~C
  Dunn, Marcus~J Hamilton, Norman Mercado-Silva, Jeffrey~C Nekola, Jordan~G
  Okie, William~H Woodruff, et~al.
\newblock Energetic limits to economic growth.
\newblock {\em BioScience}, 61(1):19--26, 2011.

\bibitem{ribeiro2019effects}
Haroldo~V Ribeiro, Diego Rybski, and J{\"u}rgen~P Kropp.
\newblock Effects of changing population or density on urban carbon dioxide
  emissions.
\newblock {\em Nature communications}, 10, 2019.

\end{thebibliography}

\bibliographystyle{unsrt}

\end{document}